\documentclass[prd,12pt,nofootinbib,preprint]{revtex4}
 
\usepackage[T1]{fontenc}
\usepackage{amsmath,amssymb}
\usepackage{relsize}
\usepackage{epsfig}
\usepackage{url}
\usepackage{subfigure}
\usepackage{graphicx}
\usepackage{grffile}
\usepackage[usenames,dvipsnames]{color}
\usepackage{slashed}
\usepackage{array}
\usepackage{multirow}
\usepackage[colorlinks,citecolor=blue]{hyperref}
\usepackage{color}
\usepackage{cancel}
\usepackage{gensymb}
\usepackage{soul}

\def\a {\alpha}

\def\l {\lambda}

\def\bar {\overline}

\def\be {\begin{equation}}
\def\ee {\end{equation}}
\def\beq {\begin{equation}}
\def\eeq {\end{equation}}
\def\bea {\begin{eqnarray}}
\def\eea {\end{eqnarray}}

\newcommand{\besub}{\begin{subequations}}
\newcommand{\eesub}{\end{subequations}}

\def\beq{\begin{equation}}
\def\eeq{\end{equation}}
\def\barr{\begin{array}}
\def\earr{\end{array}}

\begin{document}
\title{Fermi-ball in a multicomponent dark matter framework and its gravitational wave signatures}

\author{Nabarun Chakrabarty}
\email{chakrabartynabarun@gmail.com}
\affiliation{Department of Physics, Siksha Bhavana, Visva-Bharati, Santiniketan, West Bengal 731235, India}

\author{Indrani Chakraborty}
\email{indrani300888@gmail.com}
\affiliation{Department of Physics and Material Science and Engineering, Jaypee Institute of Information Technology, A-10, Sector-62, Noida 201307,
Uttar Pradesh, India}

\author{Himadri Roy}
\email{himadri027roy@gmail.com}
\affiliation{Institute of Particle Physics and Key Laboratory of Quark and Lepton Physics (MOE), Central China Normal University, Wuhan, Hubei 430079, China}

\begin{abstract} 
It has been known that under-abundant dark matter density of an inert doublet can be replenished by an additional dark matter component, say, a fermion. We find that such a scenario can lead to the formation of stable Fermi-balls through coexisting minima of the finite temperature scalar potential. More importantly, we demonstrate that the Fermi-balls contribute sizeably to the dark matter relic density. In addition, the aforesaid coexisting minima open up the possibility of a first-order phase transition. This, in turn, triggers emission of  gravitational waves that can be tested at the proposed BBO and U-DECIGO detectors. Therefore, the present study becomes a concrete setup to embed Fermi-balls in a realistic two-component dark matter model, and, to test the same using gravitational wave signatures.
\end{abstract} 
\maketitle


\section{Introduction} 

Various experimental findings such as galaxy rotation curves and gravitational lensing point towards the existence of cosmologically stable dark matter (DM)~\cite{WMAP2007,Planck2018}. While the nature of DM is under investigation, its amount in the universe is precisely known from the latest measurements by the PLANCK satellite~\cite{Planck2020}. Conceiving DM as an elementary particle calls for extending the Standard Model (SM) since there is no such candidate in the latter~\cite{Bertone2005}. In fact, it has been hypothesised that DM is a weakly interacting massive particle (WIMP)~\cite{LeeWeinberg1977} whose interaction strengths to the particles in the thermal bath is in the \emph{weak} ball-park. However, it is this sizeable interaction strengths that has cornered most minimal WIMP scenarios in view of the null results from direct detection experiments~\cite{LUX2017,PandaX2019,XENON2018,Amole2019}.

The lack of precise information on DM quantum numbers leads to the possibility that DM consists of more than one type of particle~(a partial list is \cite{Bhattacharya2013,
Bian2014,Esch2014,Bhattacharya2017,
Ahmed2018,HerreroGarcia2017,HerreroGarcia2019,
Poulin2019,Aoki2018,Aoki2017,
Elahi2019,Borah2019,Bhattacharya2020,
Bhattacharya:2019tqq,Chakrabarty:2021kmr}).
Multi-particle DM frameworks
are interesting since they predict DM-DM interaction. And while the processes driven by such interaction can contribute to the DM relic density, they have no role in 
DM-nucleon scattering thereby keeping the direct detection rates unchanged. 

There is a growing interest in studying the possibility of a first order phase transition (FOPT), and, the consequential gravitational wave (GW) spectrum in dark matter models \cite{Blinov:2015vma,Jangid:2023lny,Bandyopadhyay:2021ipw,Chakrabarty:2022yzp,Bian:2021dmp,Deng:2020dnf,Bian:2018bxr,Costa:2022oaa,Ghosh:2022fzp,Chatterjee:2022pxf,Ghosh:2020ipy,
Shibuya:2022xkj,Addazi_2018,Benincasa:2023vyp,Allahverdi:2024ofe,Banerjee:2024fam,
Biondini:2022ggt,Borah:2024lml,Borah:2023god,PhysRevD.109.095034,Borah:2022cdx,Borah:2022vsu,Borah:2021ftr,Borah:2021ocu,Liu:2017gfg}. FOPTs involve co-exiting minima, typically dubbed as the false and the true vacua, of the free energy functional. And \cite{Hong:2020est} introduced the idea that a fermionic DM candidate can get trapped inside the false
vacuum and form compact macroscopic DM candidates called Fermi-balls. The conditions that must be satisfied here are (a) a substantial mass gap of the fermion between false and true vacua compared to the phase transition temperature, (b) asymmetry in the number densities of the fermion and its antiparticle, and (c) the fermion must carry a conserved global charge $Q$. More details can be found in \cite{Hong:2020est} and are skipped here for brevity.  

In this study, we look at the possibility of stable Fermi-ball formation in a multicomponent DM setup. The inert scalar doublet scenario (see \cite{BelyaevCacciapagliaIvanov2018,
LopezHonorez2007,LopezHonorez2010,
ArhribTsaiYuanYuan2014,ChoubeyKumar2017,
LopezHonorezYaguna2011,
IlnickaKrawczykRobens2016,CaoMaRajasekaran2007,
LundstromGustafssonEdsjo2009,
GustafssonRydbeckLopezHonorez2012,
KalinowskiKotlarskiRobens2018} and the references therein), a popular WIMP setup, is extended by adding a fermion $\chi$ and a scalar $S$, both gauge singlets. The fermion interacts with the rest of the fields through the scalar $S$. Such a scenario was considered in \cite{Chakraborti:2018lso}. The fermion $\chi$ is endowed with a global $U(1)_Q$ and is thus cosmologically stable. We revisit the calculation of relic density of the resulting two-component DM setup. It is reconfirmed that $\chi$ can replenish the relic density in parameter regions where the standalone inert doublet falls under-abundant, i.e., the inert doublet \emph{desert region}. Therefore, we focus mainly on the desert region in this study. Investigating the thermally corrected scalar potential along the direction of $S$ leads to coexisting vacua in the parameter space of interest. We also compute the ensuing GW spectrum for representative benchmarks and comment on their observability in the proposed satellite based GW detectors. Apart from these, we report stable Fermi-ball formation in the desert region and compute their contribution to the observed DM abundance. In all, this study realises Fermi-balls  in an ultraviolet (UV)-complete DM model.

This paper is organized as follows. We theoretical set up is introduced in section \ref{model} and the ensuing multicomponent DM phenomenology is elucidated in section \ref{dm}. In section \ref{fopt_fb}, we shed light on the possibility of an FOPT concomitant with the DM aspects of the scenario. The GW spectrum arising out of such an FOPT is also quantified. In addition, we investigate the formation of stable Fermi-balls in the relevant parameter space of the model and estimate the contribution of Fermi-balls to the observed DM relic density. We summarise in section \ref{summary}. Important formulae are relegated to the Appendix.

\section{Theoretical setup}\label{model}

In addition to the SM-like scalar doublet $\Phi$, the scalar sector of the present setup comprises an additional doublet $\eta$ and a gauge singlet $S$. Of these, $\Phi$ and $S$ respectively pick up vacuum expectation values (VEVs) $v$ and $v_S$. On the other hand, an odd $\mathbb{Z}_2$ parity is assigned to $\eta$ in view of its DM candidacy thereby preventing the same to pick up a VEV. The particle content of the scalar multiplets is given below.
\bea
\Phi = \begin{pmatrix}
G^+ \\
\frac{1}{\sqrt2}(v + h_0 + i G_0)
\end{pmatrix},
~~~~~~\eta = \begin{pmatrix}
\eta^+   \\
\frac{1}{\sqrt2}(\eta_R + i\eta_I)
\end{pmatrix},~~~~~~~ S = v_S + s_0
\eea
With the $\mathbb{Z}_2$ ensuring stability on a cosmological time scale, the neutral inert scalars $\eta_R$ and $\eta_I$ become potential DM candidates. The scalar potential consistent with the gauge and $\mathbb{Z}_2$ symmetries reads.

\bea
V(H,\eta,S) &=& -m^2_\Phi \Phi^\dagger \Phi
 + m^2_\eta \eta^\dagger \eta
 - \frac{1}{2} m^2_{S} S^2 - \frac{\mu_S}{3}S^3 
+ \l_1 (\Phi^\dagger \Phi)^2 + \l_2 (\eta^\dagger \eta)^2 \nonumber \\
&&
+ \l_3 (\Phi^\dagger \Phi)(\eta^\dagger \eta) 
+ \l_4(\Phi^\dagger \eta) (\eta^\dagger \Phi) 
+ [\frac{1}{2}\l_5 (\eta^\dagger \Phi)^2 + h.c.]
\nonumber \\
&&
+ \frac{1}{2}\l_6 \Phi^\dagger \Phi S^2
+ \frac{1}{2}\l_7 \eta^\dagger \eta S^2 
+ \frac{1}{4}\l_8 S^4.
\eea
It is demanded that $m^2_\Phi>0,m^2_\eta>0,m^2_S>0$ in order to ensure the said configuration of the VEVs of the neutral scalars in this setup. The tadpole conditions lead to
\besub
\bea
m_\Phi^2 &=& \l_1 v^2 + \frac{1}{2}\l_6 v^2_S, \\
m_S^2 &=& -\mu_S v_S + \l_8 v_S^2 + \frac{1}{2} \l_6 v^2. 
\eea
\eesub
The $\mathbb{Z}_2$ symmetry forbids $\eta$ to mix with the other multiplets and thus the inert scalars are mass eigenstates with masses $M_{\eta_R},M_{\eta_I}$ and $M_{\eta^+}$ as expressed below.
\besub
\bea
M^2_{\eta_R} &=& m^2_\eta + \frac{1}{2}(\l_3 + \l_4 + \l_5) v^2 + \frac{1}{2}\l_7 v_S^2, \\
M^2_{\eta_I} &=& m^2_\eta + \frac{1}{2}(\l_3 + \l_4 - \l_5) v^2 + \frac{1}{2}\l_7 v_S^2, \\
M^2_{\eta^+} &=& m^2_\eta + \frac{1}{2}\l_3 v^2 + \frac{1}{2}\l_7 v_S^2,
\eea
\eesub 
On the other hand, as shown below, an $h_0$-$s_0$ mixing does take place controlled by an angle $\theta$. It leads to the mass eigenstates $h$ and $H$. 
\bea
\begin{pmatrix}
h_0 \\
s_0
\end{pmatrix} =
\begin{pmatrix}
\text{cos}\theta~~~\text{sin}\theta\\
-\text{sin}\theta~~~\text{cos}\theta
\end{pmatrix}
\begin{pmatrix}
h\\
H
\end{pmatrix}
\eea

One defines $\l_L = \l_3 + \l_4 + \l_5$ in motivated from the fact that the $\eta_R-\eta_R-h$ interaction strength in the $s_\theta \to 0$ limit is $-\l_L v$. With this,
we deem the following parameters to be independent while describing the scalar sector:
$\{M_h,M_H,M_{\eta_R},M_{\eta_I},M_{\eta^+},\mu_S,v_S,\l_L,\l_2,\l_7,s_\theta \}$. The various quartic couplings and $m_\eta$ are expressible in terms of the independent parameters as
\besub
\bea
m^2_\eta &=& M^2_{\eta_R} - \frac{1}{2}\l_L v^2 - \frac{1}{2}\l_7 v^2_S, \\
\l_1 &=& \frac{M^2_h c^2_\theta + M^2_{H} s^2_\theta}{2 v^2}, \\
\l_3 &=& \l_L + \frac{2(M^2_{\eta^+} - M^2_{\eta_R})}{v^2}, \\
\l_4 &=& \frac{(M^2_{\eta_R} + M^2_{\eta_I} - 2 M^2_{\eta^+})}{v^2}, \\
\l_5 &=& \frac{(M^2_{\eta_R} - M^2_{\eta_I})}{v^2}, \\
\l_6 &=& \frac{2(M^2_{H} - M^2_h)s_\theta c_\theta}{v v_S}, \\
\l_8 &=& \frac{M^2_h s^2_\theta + M^2_{H} c^2_\theta}{2 v_S^2} + \frac{\mu_S}{2 v_S}.
\eea
\eesub
Further, an $SU(2)_L$ singlet Dirac fermion $\chi$ is introduced that is charged under a global $U(1)_Q$. This symmetry is imposed with Fermi-ball formation in mind. Further, the same also makes $\chi$ a DM candidate by stabilising it cosmologically.   The Lagrangian involving $\chi$ therefore reads
\bea
\mathcal{L}_Y &=& -m_\chi \bar{\chi}\chi- y_\chi \bar{\chi}\chi S.
\eea
We note that the symmetries of the setup permit an inclusion of a bare mass parameter $m_\chi$. The physical mass of $\chi$ therefore becomes $M_\chi = m_\chi + y_\chi v_S$. We choose to describe the fermionic sector in terms of the independent parameters $\{M_\chi,y_\chi \}$.

\section{DM phenomenology}\label{dm}

The setup admits two DM candidates, i.e., the CP-even inert scalar $\eta_R$ and the fermion $\chi$, by virtue of the $\mathbb{Z}_2$ and $U(1)_Q$ symmetries. It is mentioned that the only way $\chi$ interacts with the SM is via the scalar $S$. The relic abundance of $\chi$ is therefore largely dictated by its annihilations to  SM particles and $H$, all  imperatively s-channel processes mediated by $h$ and $H$. As for $\eta_R$, the annihilation channels are (i) s-channel $\eta_R~\eta_R \to SM~SM$ processes mediated by $h,H$, (ii) t/u-channel $\eta_R~\eta^+~(\eta_I) \to V~V$ processes mediated by $\eta^\pm~(\eta_I)$ with $V=W^\pm Z$ as appropriate, and (iii) processes driven by the $\eta_R-\eta_R-h-h, \eta_R-\eta_R-h-H,\eta_R-\eta_R-H-H,\eta_R-\eta_R-V-V$ four-point interactions. Co-annihilation of $\eta_R$  
with the heavier inert scalars can also get triggered for small mass splittings thereby contributing to the relic density. In addition the $\chi\chi \leftrightarrow \eta_R \eta_R$ conversion processes can also a play a prominent role when there are two dark sectors involved as in the present case.

The expressions for $\chi \chi \longrightarrow \eta_R \eta_R, \eta_I \eta_I, \eta^+ \eta^-$  annihilation cross-section are given below for the sake of completion.
\besub
\bea
\sigma_{\chi \chi \to \eta_R \eta_R} &=& \frac{1}{16 \pi s} \sqrt{\frac{s - 4 M_\chi^2}{s - 4 M_{\eta_R}^2}}|\frac{y_{h \chi \chi} \l_{h\eta_R \eta_R}}{s - M_h^2 + i M_h \Gamma_h} + \frac{y_{H \chi \chi} \l_{H\eta_R \eta_R}}{s - M_H^2 + i M_H \Gamma_H}|^2 (s - 4 M_\chi^2), \\
\sigma_{\chi \chi \to \eta_I \eta_I} &=& \frac{1}{16 \pi s} \sqrt{\frac{s - 4 M_\chi^2}{s - 4 M_{\eta_I}^2}}|\frac{y_{h \chi \chi} \l_{h\eta_I \eta_I}}{s - M_h^2 + i M_h \Gamma_h} + \frac{y_{H \chi \chi} \l_{H\eta_I \eta_I}}{s - M_H^2 + i M_H \Gamma_H}|^2 (s - 4 M_\chi^2), \\
\sigma_{\chi \chi \to \eta^+ \eta^-} &=& \frac{1}{16 \pi s} \sqrt{\frac{s - 4 M_\chi^2}{s - 4 M_{\eta^+}^2}}|\frac{y_{h \chi \chi} \l_{h\eta^+ \eta^-}}{s - M_h^2 + i M_h \Gamma_h} + \frac{y_{H \chi \chi} \l_{H\eta^+ \eta^-}}{s - M_H^2 + i M_H \Gamma_H}|^2 (s - 4 M_\chi^2).
\eea
\label{sigma_conv}
\eesub
The Yukawa and trilinear couplings appearing above are expressed in the Appendix. 
The thermal relics of $\chi$ and $\eta_R$ are obtained by 
solving the coupled Boltzmann equations below. We define $x = \mu_{\text{red}}/T$, where $\mu_{\text{red}}$ denotes the reduced mass defined through $~\mu_{\text{red}}= \frac{M_{\chi}M_{\eta_R}}{M_{\chi}+M_{\eta_R}}$.
\besub
\bea
\frac{dy_{\chi}}{dx} &=& \frac{-1}{x^2}\bigg{[}\langle \sigma v_{\chi\chi \rightarrow XX}\rangle \left(y_{\chi}^{2}-(y_{\chi}^{EQ})^2\right)~+~\langle \sigma v_{\chi \chi\rightarrow \eta_R \eta_R}\rangle \left ( y_{\chi}^{2}-\frac{(y_{\chi}^{EQ})^2}{(y_{\eta_R}^{EQ})^2} y_{\eta_R}^{2}\right)\Theta(M_{\chi}-M_{\eta_R}) \nonumber \\
&&
-~\langle\sigma v_{\eta_R \eta_R \rightarrow \chi \chi}\rangle \left( y_{\eta_R}^{2}-\frac{(y_{\eta_R}^{EQ})^2}{(y_{\chi}^{EQ})^2}y_{\chi}^{2}\right)~\Theta(M_{\eta_R}-M_{\chi})\bigg{]} \\
\frac{dy_{\eta_R}}{dx} &=& \frac{-1}{x^2}\bigg{[} \langle \sigma v_{\eta_R \eta_R\rightarrow XX} \rangle \left (y_{\eta_R}^{2}-(y_{\eta_R}^{EQ})^2\right )~+~\langle \sigma v_{\eta_R\eta_R\rightarrow \chi \chi}\rangle \left (y_{H}^{2}-\frac{(y_{\eta_R}^{EQ})^2}{(y_{\chi}^{EQ})^2}y_{\chi}^{2}\right )\Theta(M_{H}-M_{1}) \nonumber \\
&&
-~\langle \sigma v_{\chi \chi \rightarrow \eta_R \eta_R}\rangle \left (y_{\chi}^{2}-\frac{(y_{\chi}^{EQ})^2}{(y_{\eta_R}^{EQ})^2}y_{\eta_R}^{2}\right )\Theta(M_{\chi}-M_{\eta_R})\bigg{]}.
\eea
\label{eq:cBEQ}
\eesub

Here $y_{i}$ ($i = \chi,\eta_R$) is related to the co-moving number density $Y_i=\frac{n_i}{s}$ (where $n_i$ refers to DM density and $s$ is entropy density) 
by $y_i=0.264M_{\text{Pl}}\sqrt{g_*}\mu Y_{i}$; similarly for equilibrium density, $y_i^{EQ}= 0.264M_{\text{Pl}}\sqrt{g_*}\mu Y_{i}^{EQ}$, with equilibrium distributions ($Y_{i}^{EQ}$) have the form
\bea
Y_{i}^{EQ}(x) = 0.145\frac{g}{g_{*}}x^{3/2}\bigg{(}\frac{m_{i}}{\mu}\bigg{)}^{3/2}e^{-x\big{(}\frac{m_{i}}{\mu}\big{)}}.
\eea 
Here, $M_{\rm Pl} = 1.22\times10^{19} ~{\rm GeV}$, $g_{*} = 106.7$\footnote{One is supposed to use $g_{*s}$ in the above equations. However, $g_{*s} \simeq g_{*}$ holds for temperatures $\sim \mathcal{O}$ (GeV) or above\cite{Kolb:1990vq}.} and $m_i$ denotes $M_\chi$ and $M_{\eta_R}$. Further, 
$X$ in eqn.~\ref{eq:cBEQ} denotes the SM particles, $\eta^{\pm}$ and $\eta_I$. This is because $\eta^{\pm}$ is expected to be in equilibrium with the thermal plasma on account of electromagnetic interactions. Also, $\eta_I$ being heavier than $\eta_R$ can decay to $\eta_R$ and SM fermions ($f$) via an off-shell $Z \to f\bar{f}$. Both $\eta^+$ and $\eta_I$ thus remain in equilibrium with the thermal bath. The thermally averaged annihilation cross section, given by 
\bea
\langle \sigma v\rangle = \frac{1}{8m^{4}_{i}T K_2^2(\frac{m_{i}}{T})}\int\limits^{\infty}_{4m_{i}^2}\sigma(s-4m_{i}^2)\sqrt{s}K_1\bigg{(}\frac{\sqrt{s}}{T}\bigg{)}ds
\label{eq:sigmav}
\eea
is evaluated at the freeze-out temperature $T_f$ and denoted by $\langle \sigma v \rangle_f$. One notes that $T_f$ is derived from 
the equality condition of DM interaction rate $\Gamma = n_{\rm DM} \langle \sigma v \rangle$ with the rate of expansion of the universe $\bar{H}(T) \simeq \sqrt{\frac{\pi^2 g_*}{90}}\frac{T^2}{M_{\rm Pl}}$. Further, $K_{1,2}(x)$ are the modified Bessel functions in eqn.(\ref{eq:sigmav}).

It must be noted that the contribution to the Boltzmann equations coming from the DM-DM conversion depends on the mass hierarchy of DM particles. Thus the use of the $\Theta$-function in the above equations. These coupled equations can be solved numerically to find the asymptotic abundance of the DM particles, $y_{i} \left (\frac{\mu_{\text{red}}}{m_{i}}x_{\infty} \right)$, which can be further used to express the contributions to relic density as
\besub
\bea
\Omega_{i}h^2 &=& \frac{854.45\times 10^{-13}}{\sqrt{g_{*}}}\frac{m_{i}}{\mu}y_{i}\left ( \frac{\mu}{m_{i}}x_{\infty}\right ),
\eea
\eesub
where $x_{\infty}$ indicates an asymptotic value of $x$ after the freeze-out. The index $i$ stands for DM components in our scenario: $\chi, \eta_R$. 
\noindent However, we use numerical techniques to solve for relic density of this two component model. The model is first implemented in \texttt{LanHEP}~\cite{Semenov:2008jy}. A compatible output was then fed into 
the publicly available tool \texttt{micrOMEGAs}4.3 ({capable of handling multi-partite DM scenarios)\cite{Belanger:2014vza} to compute the relic densities of $\chi$ and $\eta_R$.

In order to gain some insight on the relic density generation, we fix $v_S = M_H = \mu_S$ = 200 GeV, $\mu_S = 50~\text{GeV},~\l_2 = \l_L = 0.01,~\text{sin}\theta = 5 \times 10^{-4}$ 
and plot the individual relic densities versus $M_\chi$ for certain fixed values of $M_{\eta_R},y_\chi$ and $\l_7$. Fig.\ref{dm_plot_1} displays such plots corresponding to $M_{\eta_R}$ = 300 GeV, $\l_7=1,2~y_\chi=0.5,1,2$. Resonance dips noted at $M_\chi = M_h/2,~M_H/2$, and, the more the value of $y_\chi$, the lesser is $\Omega_\chi h^2$. Both these observations are expected of the $h,H$-mediated s-channel amplitudes that are also proportional to $y_\chi$. We find from Fig.\ref{dm_plot_1} that $\Omega_\chi h^2$ in the observed ballpark is obtained for $y_\chi \gtrsim 1$ for the values chosen for $v_S,M_H$ and $\mu_S$. A general feature is that $\Omega_\chi h^2$ starts increasing for $M_\chi > M_H$. A mild dependence on $\l_7$ is also noted owing to the $\chi \chi \leftrightarrow \eta_R \eta_R$ conversion amplitude being approximately proportional to $\l_7$ for small $s_\theta$.

\begin{figure*}[!htb]
\centering
\includegraphics[scale=0.45]{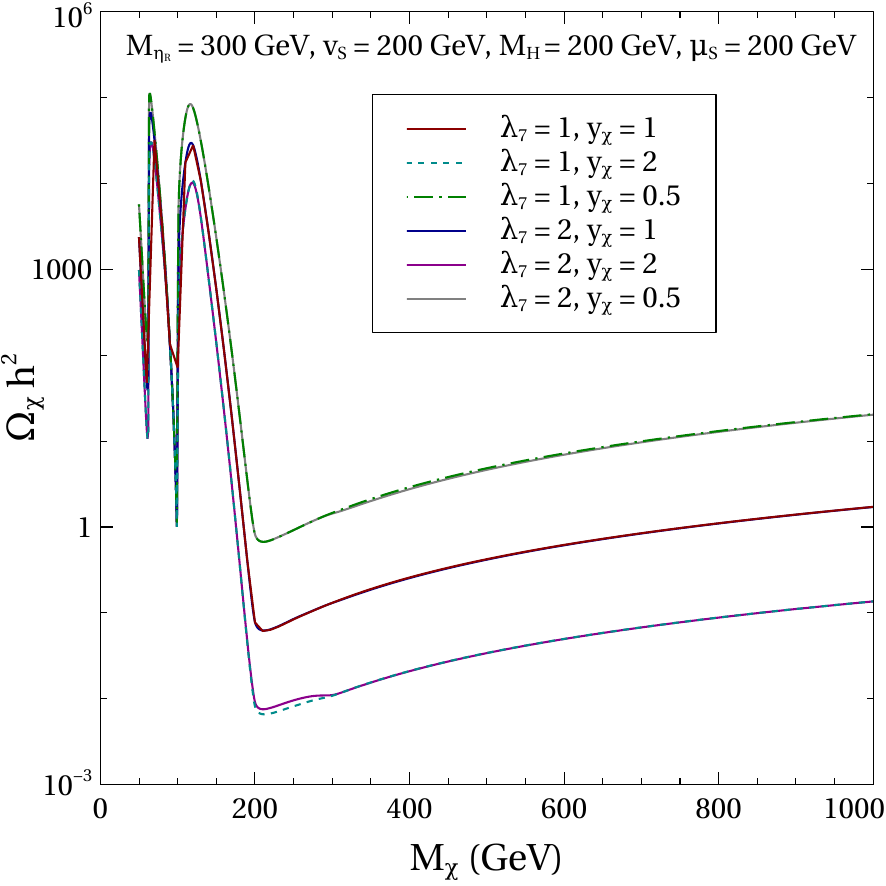}~~~
\includegraphics[scale=0.45]{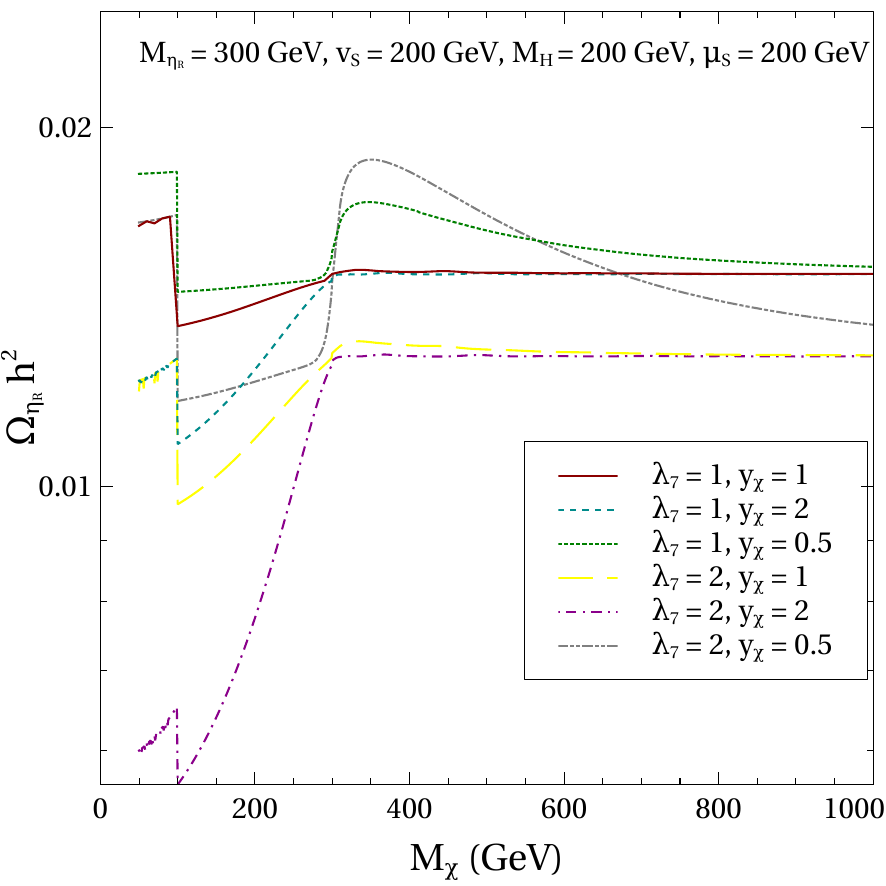}
\caption{Variation of $\Omega_\chi h^2$ (left) and $\Omega_{\eta_R} h^2$ (right) versus $M_\chi$ for $M_{\eta_R}$ = 300 GeV. The color coding is explained in the legends.}
\label{dm_plot_1}
\end{figure*}

The contribution of the inert doublet to the observed relic density is $\simeq 10\%$ in the desert region. In addition to the (co)annihilations processes for the standalone inert doublet model, $\eta_R \eta_R \to hH,HH$ become operative in the present setup. The sensitivity to $\l_7$ is seen to be higher in case of $\Omega_{\eta_R}h^2$. This is confirmed by the right plot of Fig.\ref{dm_plot_1}. The individual relic densities for $M_{\eta_R}$ = 500 GeV shown in Fig.\ref{dm_plot_2} also affirm the aforementioned observations.

\begin{figure*}[!htb]
\centering
\includegraphics[scale=0.45]{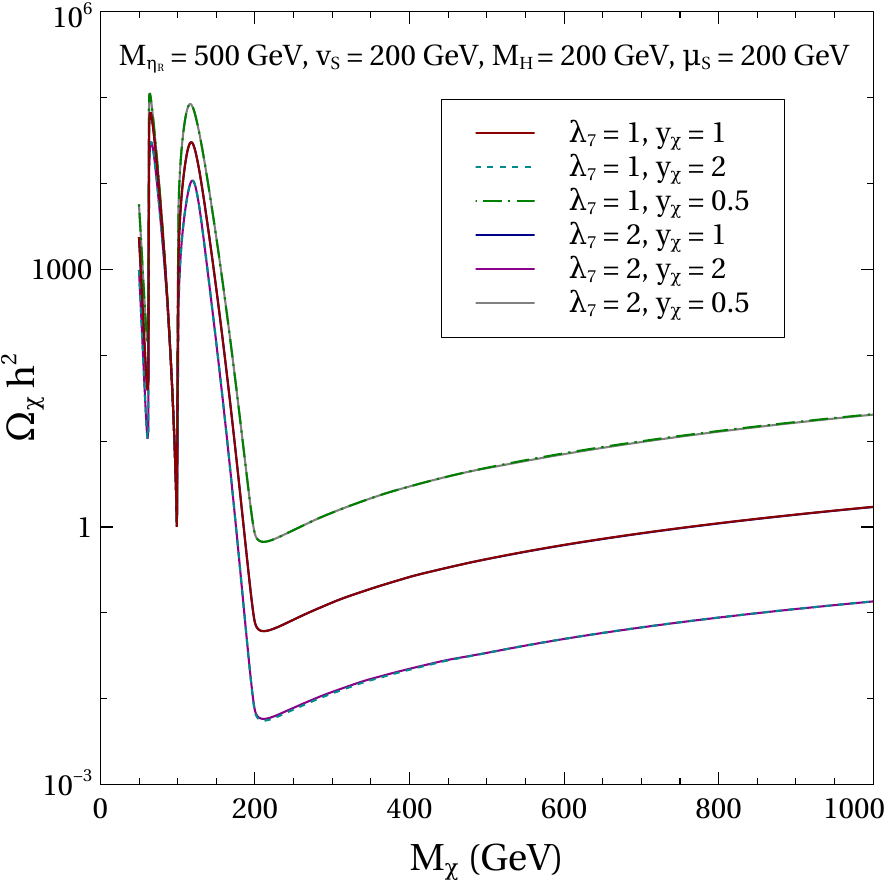}~~~
\includegraphics[scale=0.45]{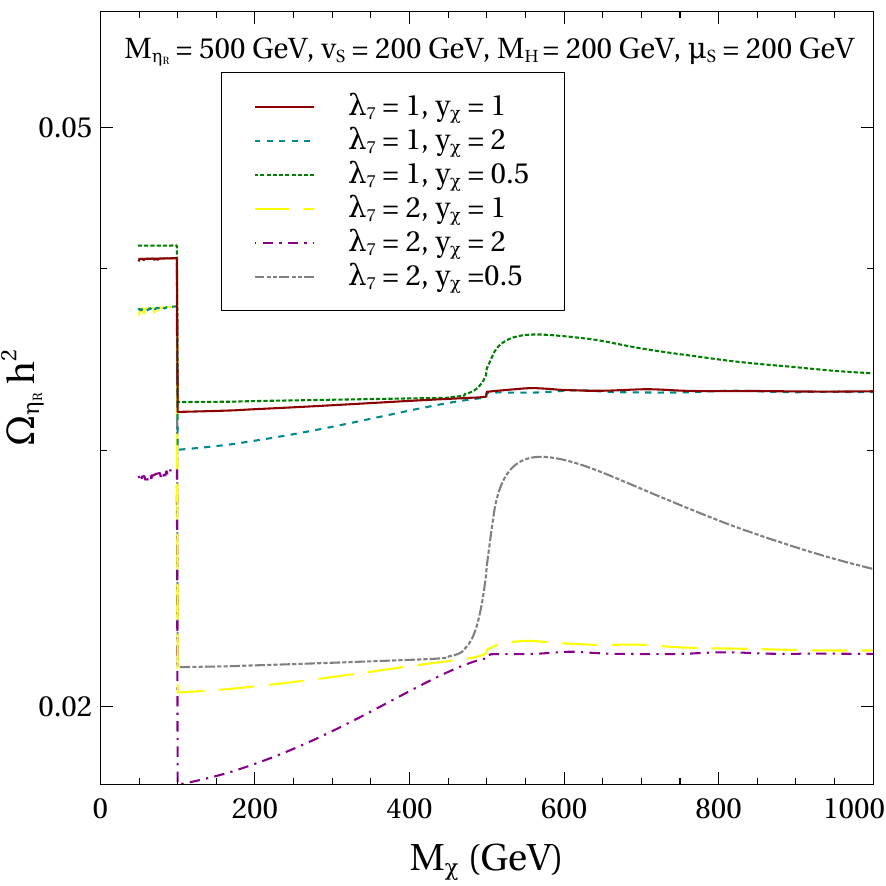}
\caption{Variation of $\Omega_\chi h^2$ (left) and $\Omega_{\eta_R} h^2$ (right) versus $M_\chi$ for $M_{\eta_R}$ = 500 GeV. The color coding is explained in the legends.}
\label{dm_plot_2}
\end{figure*}

Direct detection experiments such as LUX~\cite{LUX:2016ggv}, PandaX-II~\cite{PandaX-II:2017hlx} and Xenon-1T~\cite{XENON:2018voc} look
for DM-nucleon scatterings in terrestrial detectors. However, the non-observation of such processes has put stringent upper bounds on the corresponding cross sections. In our setup, both the DM candidates interact with the nucleons via t-channel processes mediated by $h$ and $H$. The spin-independent direct detection (SI-DD) cross sections for $\eta_R$ and $\chi$ respectively are
\besub
\bea
\sigma^{SI}_{\eta_R} = \frac{\mu_{H,n}^2}{4\pi}\bigg{[}\frac{m_n~f_n}{M_{\eta_R}~v}\bigg{(}\frac{\l_{h\eta_R\eta_R}}{M_h^2}+\frac{\l_{H\eta_R\eta_R}}{M_H^2}\bigg{)}\bigg{]}^2.\\
\sigma^{SI}_{\chi} = \sin{2\theta}\frac{\mu_{N_1,n}^2}{4\pi}\bigg{[}\frac{y_{\chi}~m_n~f_n}{v}\bigg{(}\frac{1}{M_H^2}-\frac{1}{M_h^2}\bigg{)}\bigg{]}^2
\eea
\eesub
where $\mu_{\eta_R,n} = m_n M_{\eta_R} /(m_n + M_{\eta_R})$, $\mu_{\chi,n} = m_n M_\chi /(m_n + M_\chi )$ are the DM-nucleon reduced masses and $f_n$ = 0.2837 is the nucleon form factor~\cite{Alarcon:2012nr}. In this two-component DM framework, the 
\emph{effective} SI-DD cross sections relevant for each of the candidates can be expressed by the individual DM-nucleon cross-section scaled by the relative abundance of that particular component ($\Omega_i h^2$) in the observed DM relic density ($\Omega_{\text{obs}} h^2$). That is,
\bea
\sigma^{SI}_{i,eff} = \frac{\Omega_i h^2}{\Omega_{\text{obs}}h^2}~\sigma_{i}^{SI}.
\eea
We adopt $\Omega_{\text{obs}}h^2$ = 0.12 here.
It is ensured in the subsequent analysis that $\sigma^{SI}_{\eta_R,eff}$ and $\sigma^{SI}_{\chi,eff}$ respect the latest XENON-1T bounds. A more careful analysis for multi-particle DM direct search cross section can be performed by computing total recoil rate (see for instance, \cite{Bhattacharya2017,Ahmed2018,
HerreroGarcia2017,HerreroGarcia2019}), however the above procedure 
gives a correct order of magnitude estimate for individual components.

\section{Dynamics of a FOPT}\label{fopt_fb}

Since $\chi$ couples only to the scalar $S$ prior to EWSB, the Fermi-ball dynamics is dictated by the thermal evolution of the vacuum expectation value $<S(T)>$. Therefore, 
we examine the scalar potential as a function of a background field $\phi$ in the direction of the scalar $S$. The tree level potential is straightforwardly found to be
\bea
V_0(\phi) &=& -\frac{1}{2}m^2_S \phi^2 - \frac{1}{3}\mu_S \phi^3 + \frac{1}{4}\l_8 \phi^4.
\eea
Next, we quote the corresponding one-loop Coleman-Weinberg potential~\cite{ColemanWeinberg1973} in dimensional regularisation (DR). That is,
\bea
V_{\text{CW}}(\phi) &=& \frac{1}{64 \pi^2}\sum_{i} n_i \bigg\{M^4_i(\phi)~\text{log}\bigg(\frac{M^2_i(\phi)}{\mu^2} \bigg)          - \frac{3}{2}\bigg\}.
\eea
Here, $\mu$ is the renormalisation scale encountered in dimensional regularisation (DR) and $i$ runs over all particles interacting with $S$. The $i$th particle has $n_i$ number of degrees of freedom, and, picks up a field dependent mass $M_i(\phi)$ in the process. One finds 
\bea
n_{G^+} = n_{\eta^+} = 2,~n_{G_0} = n_{h_0} = n_{\eta_R} = n_{\eta_I} = 1,~n_\chi = -4.
\eea
The various field-dependent masses are expressed as
\besub
\bea
M^2_{G^+}(\phi) &=& M^2_{G_0}(\phi)= M^2_{h_0}(\phi) = -m_\Phi^2 + \frac{1}{2}\l_6 \phi^2, \\
M^2_{\eta^+}(\phi) &=& M^2_{\eta_I}(\phi)= M^2_{\eta_R}(\phi) = m_\eta^2 + \frac{1}{2}\l_7 \phi^2, \\
M^2_S(\phi) &=& -m^2_S - 2\mu_S \phi + 3\l_8 \phi^2,\\
M^2_\chi(\phi) &=& (m_f + y_\chi \phi)^2.
\eea
\eesub
We adopt the $\overline{\text{MS}}$ scheme  in this study and thus subsequently choose $\mu = v_S$. Further, the one-loop correction to the scalar potential induced due to $T \neq 0$ reads~\cite{DolanJackiw1974,Quiros:1999jp,Laine:2016hma}
\bea
V_T(\phi,T) &=& \frac{T^4}{2\pi^2}\Bigg[\sum_{b=\text{boson}} n_b J_B\bigg(\frac{M^2_b(\phi)}{T^2}\bigg) + \sum_{f=\text{fermion}} n_f J_F\bigg(\frac{M^2_f(\phi)}{T^2}\bigg)\Bigg].
\eea
In the above, the indices $b$ and $f$ respectively run over the bosonic and fermionic fields. The functions $J_{B,F}(x)$ read
\bea
J_{B,F}(x) &=& \int_0^\infty dx~y^2 ~\text{log}[1 \mp e^{-\sqrt{y^2 + x}}].
\eea 
Further, infrared effects are included using the daisy remmuation technique~\cite{Parwani1992,ArnoldEspinosa1993}. In particular, the Arnold-Espinosa prescription~\cite{ArnoldEspinosa1993} that contributes the following term:
\bea
V_{\text{Daisy}}({\phi,T}) &=& -\frac{T}{12\pi}  \sum_{b=\text{boson}} n_b\bigg[M^3_b(\phi,T) - M^3_b(\phi)\bigg], 
\eea
where $M^2_b({\phi,T}) = M^2_b({\phi}) + \Pi_b(T)$ refers to the thermal mass of the $b$th boson. The Debye mass corrections $\Pi_b(T)$ for the considered setup are relegated to the Appendix. In all, the scalar potential finite temperature becomes
\bea
V_{\text{total}}(\phi,T) &=& V_0(\phi) + V_{\text{CW}}(\phi) + V_{\text{Daisy}}(\phi,T).
\eea
The scalar potential above admits coexisting minima that can be obtained through
\besub
\bea
\frac{\partial V_{\text{total}}}{\partial \phi} = 0, \\
\frac{\partial^2 V_{\text{total}}}{\partial \phi^2} > 0.
\eea
\eesub
The two minima $\phi_f$ and $\phi_t$ (say) can be dubbed as the false and the true vacuum respectively. Such dynamics therefore
opens up the possibility of a first order phase transition (FOPT). It also entails tunnelling from $\phi_f$ to $\phi_t$ whose probability per unit 4-volume is given by~\cite{Linde1983}
\bea
\Gamma(T) &=& T^4 \bigg(\frac{S_E}{2\pi T}\bigg)^{3/2}e^{-\frac{S_E}{T}}.
\eea
Here, $S_E$ is the classical euclidean "bounce" action in 3-dimensions calculated as
\bea
S_E &=& 4\pi \int_0^\infty dr~r^2 \bigg[\frac{1}{2}\Big( \frac{d \phi}{d r} \Big)^2 + V_{\text{total}}(\phi,T) \bigg].
\eea
The scalar field $\phi$ is derived by solving the classical field equation
\bea
\frac{d^2 \phi}{d r^2} + \frac{2}{r}\frac{d\phi}{d r} - \frac{\partial V_{\text{total}}}{\partial \phi} = 0.
\eea

A critical temperature $T_c$ is identified through
\bea
V_{\text{total}}(\phi_f(T_c),T_c) = V_{\text{total}}(\phi_t(T_c),T_c),
\eea
The bubble picture can be invoked to understand FOPT in analogy with the liquid-gas phase transition. In a sea of the false vacuum can be thought to be populated by bubbles containing the true vacuum. And the bubbles grow in size as tunnelling from $\phi_f$ to $\phi_t$ proceeds. There is $\sim$ 1 bubble per unit Hubble volume at the nucleation temperature $T_n$ typically defined as
\bea
\frac{S_E(T_n)}{T_n} = 140.
\eea

\begin{table}[htpb]
\centering
\begin{tabular}{ |c | c | c| } 
\hline
  & BM1 & BM2  \\  \hline
$M_{\eta_R}$ & 351.194 GeV & 228.89 GeV \\  
$M_{\chi}$ & 602.952 GeV & 618.55 GeV \\  
$y_{\chi}$ & 1.754 & 1.838 \\  
$\Omega_{\eta_R}h^2$ & 0.019 & 0.010 \\  
$\Omega_{\chi}h^2$ & 0.081 & 0.072 \\
$\sigma^{\text{SI}}_{\eta_R}$ & 6.936 $\times 10^{-48}~\text{cm}^2$ & 1.62 $\times 10^{-47}$ cm$^2$ \\ 
$\sigma^{\text{SI}}_{\chi}$ & 8.25 $\times 10^{-50}$ cm$^2$ & 8.67 $\times 10^{-50}$ cm$^2$\\   
$T_c$ & 263.784 GeV & 380.746 GeV \\
$\frac{\phi_c}{T_c}$ & 1.491 & 2.018\\  
$T_n$ & 236.2 GeV & 284.2 GeV\\
$\alpha_n$ & $1.23 \times 10^{-2}$ & 1.84 $\times 10^{-3}$\\
$\beta$ & 2.219 $\times 10^{3}$ & 1.025 $\times 10^{3}$\\
$v_b$ & 0.660 & 0.611\\
$\frac{M^*_\chi}{T_*}$ & 3.805 & 4.505\\
$F_\chi$ & 0.459 & 0.590\\ 
$\Omega_{\text{FB}}h^2/\Omega_{\text{obs}}h^2$ & 16.67$\%$ & 31.67$\%$\\ 
Required $c_\chi$ & $4.93 \times 10^{-3}$  & $7.29 \times 10^{-3}$ \\ \hline
\end{tabular}
\caption{Benchmark parameter points and the corresponding predictions of GW amplitude and Fermi-ball contribution to DM.}
\label{bp}
\end{table}
Gravitational waves in case of an FOPT are mainly generated through (a) bubble wall collisions \cite{TurnerWilczek1990,Kosowsky:1991ua,
Kosowsky:1992rz,KosowskyTurner1993,PhysRevD.46.2384}, (b) sound waves \cite{Hindmarsh:2013xza,Giblin2014,Hindmarsh:2015qta,
Hindmarsh:2017gnf}, and (c) 
magneto-hydrodynamic (MHD) turbulence \cite{Kosowsky2002,Caprini2006,
Gogoberidze2007,Caprini2009,Niksa2018} in the plasma. We express below the corresponding contributions to the energy density as a function of the GW frequency $f$ \cite{Caprini2016,Guo2021,Caprini2020,
Kamionkowski1994,Steinhardt1982,Espinosa2010}.
\besub
\bea
\Omega_{\text{coll}}h^2 &=&  1.67 \times 10^{-5} \Big( \frac{\beta}{H} \Big)^{-2}\Big(\frac{0.11 v^3_b}{0.42 + v^2_b}\Big)
\Big( \frac{\kappa_c \a}{1 + \a} \Big)^2 
\Big(\frac{100}{g_*}\Big)^{\frac{1}{3}} 
\Big( \frac{3.8 (f/f_{\text{coll}})^{2.8}}{1+2.8(f/f_{\text{coll}})^{3.8}} \Big),\\
\Omega_{\text{sw}}h^2 &=&  2.65 \times 10^{-6} \Big( \frac{\beta}{H} \Big)^{-1} v_b
\Big( \frac{\kappa_s \a}{1 + \a} \Big)^2  
\Big(\frac{100}{g_*}\Big)^{\frac{1}{3}} 
\Big(\frac{f}{f_{\text{sw}}} \Big)^3
\Big( \frac{7}{4+3(f/f_{\text{sw}})^{2}} \Big)^{7/2},\\
\Omega_{\text{tur}}h^2 &=&  3.35 \times 10^{-4} \Big( \frac{\beta}{H} \Big)^{-1} v_b
\Big( \frac{\kappa_t \a}{1 + \a} 
 \Big)^{3/2} 
\Big(\frac{100}{g_*}\Big)^{\frac{1}{3}} 
\Big(\frac{f}{f_{\text{tur}}} \Big)^3
\frac{(1 + f/f_{\text{tur}})^{-11/3}}{1+ 8\pi f/h_s}. 
\eea
\eesub
In the above, the parameter $\a$~\cite{KamionkowskiKosowskyTurner1994} is defined as
\bea
\alpha = \frac{\Delta \rho}{\rho_r(T_n)},
\eea
where $\Delta \rho = \\
\bigg[V_{\text{total}}(\phi_f(T),T)-V_{\text{total}}(\phi_t(T),T) - T \frac{d}{d T}\big(V_{\text{total}}(\phi_f(T),T)-V_{\text{total}}(\phi_t(T),T) \big)\bigg]_{T=T_n}$ parameterizes the energy budget of the FOPT through its latent heat release. Moreover, $\rho_r(T) = \frac{\pi^2}{30}g_* T^4$ is the energy density of the radiation dominated phase in the presence of $g_*$ number of relativistic degrees of freedom.
One also has
\bea
\beta &=& T_n \frac{d S_E}{d T}\bigg|_{T=T_n}
\eea~\cite{Nicolis2004}
that quantifies the speed of phase transition. Further, $v_b$ is the bubble wall velocity in general related to the Jouget velocity $v_J = \frac{1/\sqrt{3} + \sqrt{\alpha^2 + 2\alpha/3}}{1+\alpha}$ \cite{Kamionkowski1994,Steinhardt1982,Espinosa2010}. One reckons $v_b \simeq v_J$ unless there is supercooling. The factors $\kappa_c,\kappa_s$ and $\kappa_t$ respectively are efficiency factors relevant to bubble collision, sound wave emission and turbulence. They are expressed as
\besub
\bea
\kappa_c &=& \frac{1}{1+0.715 \alpha}\bigg( 0.715\alpha + \frac{4}{27}\sqrt{\frac{3\alpha}{2}}\bigg),\\
\kappa_s &=& \frac{\sqrt{\alpha}}{0.135+\sqrt{0.98+\alpha}},\\
\kappa_t &=& 0.1\kappa_s.
\eea
\eesub
Finally, the frequencies $f_{coll},~f_{sw}$ and $f_{turb}$ at which the corresponding GW amplitudes peak \cite{Caprini2016,Guo2021,Caprini2020,
Kamionkowski1994,Steinhardt1982,Espinosa2010} are expressed below.
\besub
\bea
f_{\text{coll}} &=& 1.65 \times 10^{-5} \text{Hz}~\Big(\frac{g_*}{100}\Big)^{1/6}~\Big( \frac{T_n}{100~\text{GeV}} \Big) ~\Big(\frac{0.62}{v^2_b - 0.1 v_b + 1.8}\Big)\Big(\frac{\beta}{H}\Big), \\
f_{\text{sw}} &=& 1.65 \times 10^{-5} \text{Hz}~\Big(\frac{g_*}{100}\Big)^{1/6}~\Big( \frac{T_n}{100~\text{GeV}} \Big) ~\Big(\frac{2}{\sqrt3}\Big)\Big(\frac{\beta}{H}\Big), \\
f_{\text{tur}} &=& 1.65 \times 10^{-5} \text{Hz}~\Big(\frac{g_*}{100}\Big)^{1/6}~\Big( \frac{T_n}{100~\text{GeV}} \Big) ~\Big(\frac{7}{4}\Big)\Big(\frac{\beta}{H}\Big).
\eea
\eesub

The total GW density is therefore the sum of the individual contributions from bubble collsion, sound wave production and turbulence. Thus,
\bea
\Omega_{\text{GW}}h^2 =\Omega_{\text{coll}}h^2 + \Omega_{\text{sw}}h^2 + \Omega_{\text{tur}}h^2. 
\eea 

We fix $M_H = v_S = \mu_S$ = 200 GeV, $\l_L = 0.01,~s_\theta = 5 \times 10^{-4},~y_\chi = 1.8,~\l_7 = 1,2$ for illustration and make the following variation 100 GeV $\leq M_{\eta_R},M_\chi \leq$ 1 TeV in our numerical scans. The scan range is motivated from the fact that we wish to probe the IDM desert region in terms of the GW spectrum and Fermi-ball formation. The parameter regions in the $M_{\eta_R}-M_\chi$ plane corresponding to 
$\frac{\phi_c}{T_c} > 1,2,3$ are overlayed on the region accounting for $\geq 50\%$ of the observed relic abundance in Fig.\ref{scan_DM_PT}.  
\begin{figure*}[!htb]
\centering
\includegraphics[scale=0.60]{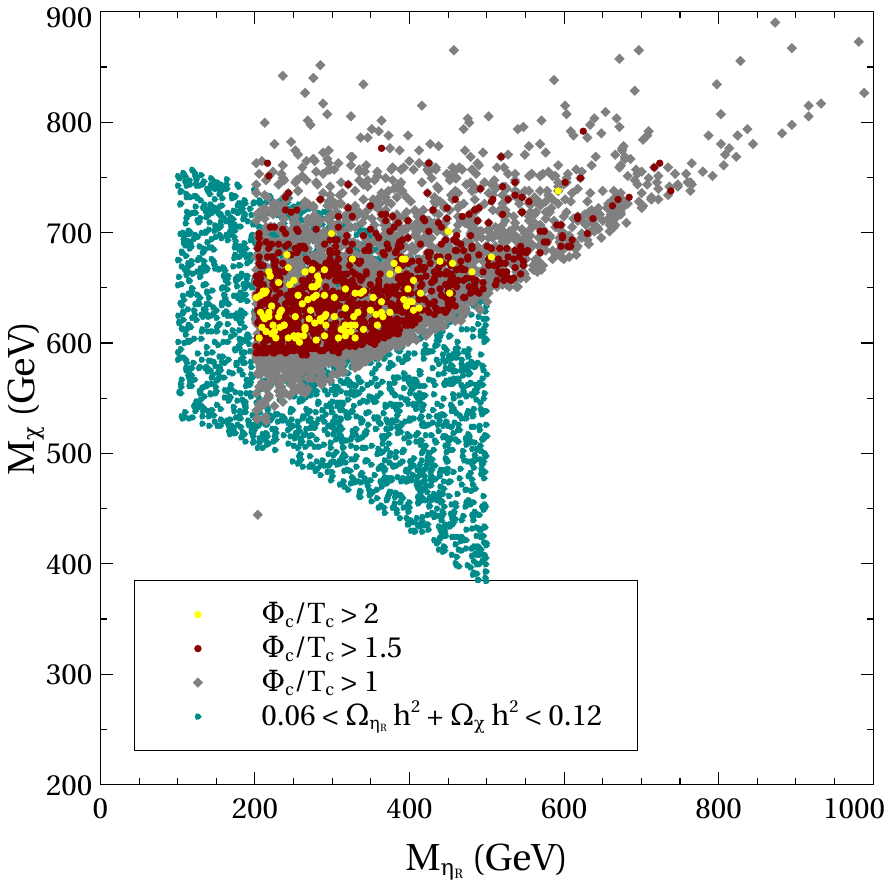}
\caption{Variation of $V_{\text{total}}(\phi,T)$ versus $\phi$ for $T=T_c,~T_n$ in case of BP1 and BP2. The color coding is explained in the legends.}
\label{scan_DM_PT}
\end{figure*}
An FOPT with $\frac{\phi_c}{T_c} \geq 1$ is dubbed \emph{strong} hereafter. The results of 
Fig.\ref{scan_DM_PT} demonstrate the possibility of accommodating together a strong FOPT and an elevated thermal relic abundance in the IDM desert region. This point emerges as a major upshot of this study. We would like to mention that the wide [0.06,0.12] interval for the thermal relic is chosen deliberately in order to leave room for a possible Fermi-ball contribution.

To advance the discussion further, we pick two 
two representative benchmarks from the desert region that predict strong FOPT. The benchmarks, BM1 and BM2, are displayed in in Table~\ref{bp}. The shape of $V_{\text{total}}(\phi,T)$ is shown at the critical and nucleation temperatures for the chosen benchmarks in Fig.\ref{VphiT_BP}.
\begin{figure*}[!htb]
\centering
\includegraphics[scale=0.45]{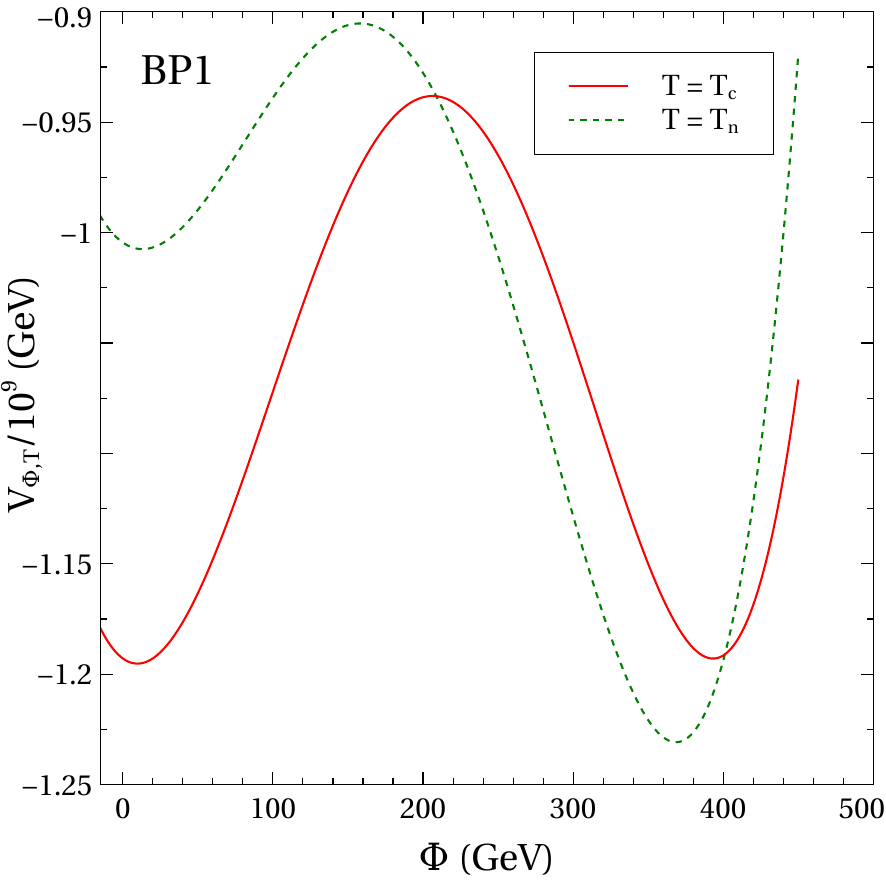}~~~
\includegraphics[scale=0.45]{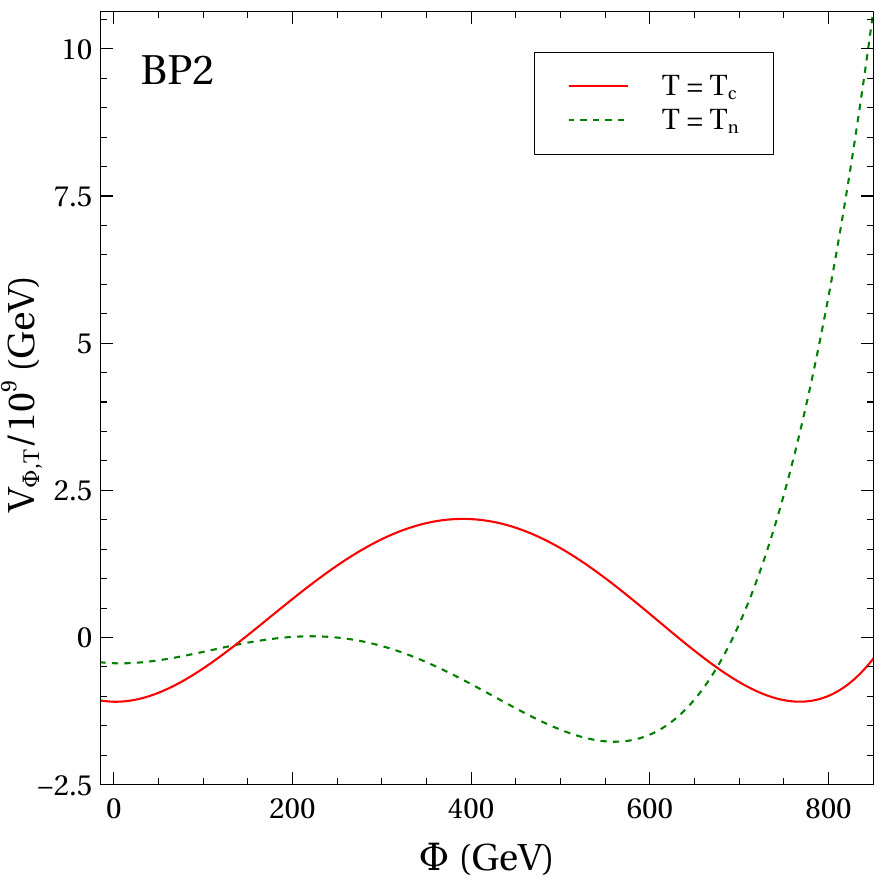}
\caption{Regions in the $M_{\eta_R}-M_{\chi}$ plane simultaneously accounting for strong FOPT and at least $50\%$ of the observed DM relic. The color coding is explained in the legends.}
\label{VphiT_BP}
\end{figure*}

The WIMP DM density for both benchmarks is majorly generated by the fermion $\chi$ though the IDM shares $\gtrsim 15\%$ of the WIMP contribution. The corresponding $S_E$ at various $T$ are computed using the publicly available tool \texttt{FindBounce} \cite{Guada:2020xnz}. The parameters relevant to an $\Omega_{\text{GW}}h^2$ calculation are also shown in the table. The GW spectra corresponding to these benchmarks can be seen in Fig.\ref{omega_GW}.
The GW spectra peak at around $\mathcal{O}(10^{-15})$ and $\mathcal{O}(10^{-17})$ values for BM1 and BM2 respectively. And the shapes of the spectra are such that BM1 is within the reach of the proposed GW detector BBO \cite{YagiSeto2011}. BM2 can also be probed by the U-DECIGO detector \cite{Sato2017,Ishikawa2021}. Given the representative nature of the chosen benchmarks, it is inferred that the IDM desert region can be probed by the aforementioned experiments.

\begin{figure*}[!htb]
\centering
\includegraphics[scale=0.45]{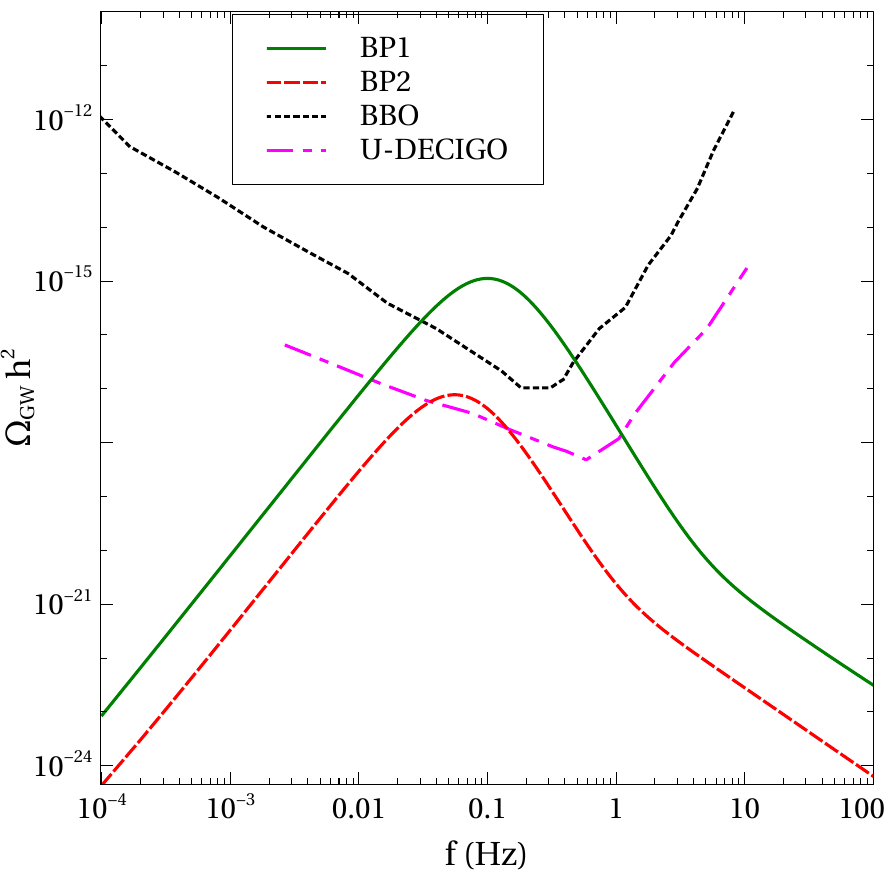}
\caption{Variation of $V_{\text{total}}(\phi,T)$ versus $\phi$ for $T=T_c,~T_n$ in case of BP1 and BP2. The color coding is explained in the legends.}
\label{omega_GW}
\end{figure*}

The possibility of Fermi-ball formation in this specific framework is discussed next. We first define $\Delta U(T) = V_{\text{total}}(\phi_f(T),T)-V_{\text{total}}(\phi_t(T),T)$.
The energy of a Fermi-ball for $T \simeq 0$ with global charge $Q_{\text{FB}}$ and radius $R$ reads
\bea
E= \frac{3\pi}{4}\bigg(\frac{3}{2\pi}\bigg)^{2/3}\frac{Q_{\text{FB}}^{4/3}}{R} + 4\pi \sigma_0 R^2 + \frac{4\pi}{3}U_0 R^3.
\eea
where the first term is the Fermi-gas pressure  of $\chi$, $\sigma_0$ the surface tension, and $U_0 = \Delta U(T)|_{T=0}$. The surface term can be neglected compared to the volume term given the macroscopic size of the Fermi-balls. 
Following closely the analysis of \cite{Hong:2020est}, we find that stability of the a Fermi-ball requires
\bea
(12\pi^2 U_0)^{1/4} < m_\chi + y_\chi \phi_t(0).
\eea
The expressions for the mass, radius and charge of a Fermi-ball can be found in \cite{Hong:2020est} and are therefore skipped here.
It is clear that Fermi-ball formation necessitates coexisting minima of the scalar potential at zero temperature. This can be arranged through the trilinear $-\frac{1}{3}\mu_S \phi^3$ term. Thus, in retrospect, the introduction of such a term stands justified.

A important quantity in the context of Fermi-balls is the fraction of $\chi$ trapped in the false vacuum. Denoting the same by $F_\chi$, it can be obtained as a function of the bubble wall velocity $v_b$ and $M^*_\chi/T_*$. Here, $T_*$ refers to the temperature where Fermi-balls start to form and $M^*_\chi$ is the field dependent mass of the fermion at that temperature. We take $T_* \simeq T_n$ in this analysis, an approximation that remains valid for a non-supercooled FOPT such as the ones under consideration. Finally, Fermi-balls contribute to the relic density by  
\bea
\Omega_{\text{FB}}h^2 = 0.12 \times F_\chi \bigg( \frac{c_\chi}{0.0146} \bigg) \bigg( \frac{U_0^{1/4}}{100~\text{GeV}}\bigg),
\eea
where $c_\chi$ is a number typically $\sim 0.01$. The observed relic is thus a sum of the contributions from DM scattering and Fermi-balls. That is,
\bea
\Omega_{\text{obs}} h^2 &=& \Omega_{\eta_R}h^2 + \Omega_{\chi}h^2 + \Omega_{\text{FB}}h^2.
\eea
The values of $F_\chi$ and the relative contributions of Fermi-balls to the observed relic density are shown in Table~\ref{bp}. It is seen that the contributions are sizeable and exceed the contibution from $\eta$. In fact, Fermi-balls account for $\simeq 32\%$ of the observed DM for BP2. In the last row of the same table, we also estimate the value of $c_\chi$ stipulated in the process.

\section{Summary}\label{summary}

We have revisited a two-component DM model involving an inert scalar $\eta$ and an  fermion $\chi$ that are respectively doublet and singlet under $SU(2)_L$. The fermion interacts with the rest of the fields through an $SU(2)_L$ scalar $S$, and, makes up for the relic density in regions where the standalone inert doublet would lead to under-abundance. This is particularly true for the 100 - 500 GeV mass range of the inert doublet where we focus on in this study.

The presence of a cubic term in the scalar potential leads to co-exiting minima. This possibility is enhanced by $T \neq 0$ corrections thereby triggering first order phase transitions. We have demonstrated in this study that the aforementioned mass range of the inert doublet can lead to a \emph{strong} FOPT for appropriate values of the other parameters. Two representative benchmarks are chosen with $T_n$ in the 200-300 GeV range. The gravitational wave spectrum arising out of FOPT is also looked at. We obtain $\Omega_{\text{GW}}h^2 \simeq \mathcal{O}(10^{-17})$ for frequency $\sim \mathcal{O}(0.1)$ Hz implying that the proposed benchmarks can be probed by the proposed BBO and U-DECIGO GW detectors.  

We have also shown that the present scenario allows for the formation of stable Fermi-balls
in the parameter region sensitive to the aforesaid detectors in terms of the strength of GW production. The contribution of Fermi-balls to DM relic density is also estimated. It is seen that the Fermi-ball contribution can be a sizeable $\gtrsim 30\%$ of the observed relic density.

\section*{Acknowledegement}

IC acknowledges support from Department of
Science and Technology, Govt. of India, under grant number IFA18-PH214 (INSPIRE Faculty Award). NC acknowledges support from
Department of Science and Technology, Govt. of India, under grant
number IFA19-PH237 (INSPIRE Faculty Award). HR is supported by the National Natural Science Foundation of China under Grant Nos. 12475094, 12135006 and 12075097, as well as by the Fundamental Research Funds for the Central Universities under Grant No. CCNU24AI003.

\section{Appendix}

\subsection{Model couplings}
\besub
\bea
y_{h\chi\chi} &=& -y_\chi s_\theta, \\
y_{H\chi\chi} &=& y_\chi c_\theta,\\
\l_{h\eta_R\eta_R} &=& \l_L v c_\theta - \l_7 v_S s_\theta,\\
\l_{h\eta_I\eta_I} &=& (\l_3 + \l_4 - \l_5) v c_\theta - \l_7 v_S s_\theta,\\
\l_{h\eta^+\eta^-} &=& \l_3 v c_\theta - \l_7 v_S s_\theta,\\
\l_{H\eta_R\eta_R} &=& \l_L v s_\theta + \l_7 v_S c_\theta,\\
\l_{H\eta_I\eta_I} &=& (\l_3 + \l_4 - \l_5) v s_\theta + \l_7 v_S c_\theta,\\
\l_{H\eta^+\eta^-} &=& \l_3 v s_\theta + \l_7 v_S c_\theta.
\eea
\eesub

\subsection{Debye mass corrections}
\besub
\bea
\Pi_{h_0}(T) &=&  \frac{1}{12}\Big(6\l_1 + 2 \l_3 + \l_4 + \frac{1}{2}\l_6 + \frac{3}{4}(g^\prime)^2 + \frac{9}{4}g^2 + 3 y^2_t \Big)T^2, \\
\Pi_{G_0}(T) &=& \Pi_{G^+}(T) = \Pi_{h_0}(T), \\
\Pi_{\eta_R}(T) &=& \frac{1}{12}\Big(6\l_2 + 2 \l_3 + \l_4 + \frac{1}{2}\l_7 + \frac{3}{4}(g^\prime)^2 + \frac{9}{4}g^2 \Big)T^2, \\
\Pi_{\eta_I}(T) &=& \Pi_{\eta^+}(T) = \Pi_{\eta_R}(T), \\
\Pi_{s_0}(T) &=& \Big(2\l_6 + 2\l_7 + 3\l_8 + 2 y^2_\chi \Big)\frac{T^2}{12}.
\eea
\eesub

\bibliographystyle{JHEP}
\bibliography{refFB} 
\end{document}